\documentclass[conference]{IEEEtran}
\usepackage{cite}
\usepackage{graphicx}
\usepackage{braket}
\usepackage{fancyhdr}

\def\zs{\mbox{{$z_{\rm spec}$}}}
\def\zp{\mbox{{$z_{\rm phot}$}}}
\begin{document}

%
\title{Probability density estimation of photometric redshifts based on machine learning}

\author{\IEEEauthorblockN{Stefano Cavuoti\\ and Massimo Brescia}
\IEEEauthorblockA{Osservatorio  di Capodimonte\\
INAF - via Moiariello 16\\
80131 - Napoli, Italy\\
Email: cavuoti@na.astro.it}
\and
\IEEEauthorblockN{Valeria Amaro\\ Civita Vellucci\\ and Giuseppe Longo }
\IEEEauthorblockA{Dipartimento di Fisica\\
Universit\'a Federico II \\
via Cintia 6, 80125 - Napoli, Italia\\
Email: amaro@na.infn.it}
\and
\IEEEauthorblockN{Crescenzo Tortora}
\IEEEauthorblockA{Kapteyn  Institute\\
University of Groningen\\
P.O. Box 800\\9700 AV Groningen, The Netherlands}}


%

\IEEEspecialpapernotice{(Invited Paper for 2016 IEEE Symposium Series on Computational Intelligence) }

\maketitle

\begin{abstract}
Photometric redshifts (photo-z's) provide an alternative way to estimate the distances of large samples of galaxies
and are therefore crucial to a large variety of cosmological problems. Among the various methods proposed over the years, supervised machine learning (ML) methods capable to interpolate the knowledge gained by means of spectroscopical data  have proven to be very effective. METAPHOR (Machine-learning Estimation Tool for Accurate PHOtometric Redshifts) is a novel method designed to provide a reliable PDF (Probability density Function) of the error distribution of photometric redshifts predicted by ML methods. The method is implemented as a modular workflow, whose internal engine for photo-z estimation makes use of the MLPQNA neural network (Multi Layer Perceptron with Quasi Newton learning rule), with the possibility to easily replace the specific machine learning model chosen to predict photo-z's.
After a short description of the software, we present a summary of results on public galaxy data (Sloan Digital Sky Survey - Data Release 9) and a comparison with a completely different method based on Spectral Energy Distribution (SED) template fitting.
\end{abstract}

%
\IEEEpeerreviewmaketitle


\section{Introduction}
Redshifts are used to constrain dark matter and dark energy contents of the Universe  \cite{serjeant2014}, to reconstruct the Universe large scale structure \cite{aragon2015}, to identify galaxy clusters
and groups \cite{capozzi2009,annunziatella2016}, to map the galaxy color-redshift relationships \cite{masters2015}, to classify astronomical sources \cite{brescia2}, to quote just a few applications.
Spectroscopic methods provide the most accurate measure of redshifts but
due to technical limitations they cannot cope with the large samples (tens of millions of objects) which are required
by modern precision cosmology.
For this reason, another approach has been increasingly used: to obtain photometric redshifts (hereafter photo-z's) by exploiting broad or medium-band imaging data.
A  plethora of methods and techniques  for photo-z's have in fact been implemented and tested on a large variety
of all-sky multi-band surveys.
These methods are based either on fitting the observed Spectral Energy Distributions (SEDs) \cite{bolzonella2000,arnouts1999,ilbert2006,tanaka2015}  to a selected library of spectral templates, or on the empirical explorations of the photometric parameter space (mostly fluxes and derived colors) aimed at learning -through interpolation- the hidden function mapping the photometric data onto the spectroscopic redshift distribution. To do so, empirical methods need a knowledge base (KB), i.e. a relatively large subset of objects for which the spectroscopic redshift is known in advance  \cite{Cavuoti+15_KIDS_I,brescia2014,carrasco2013,Connolly}.

These two approaches present complementary aspects.
SED fitting methods are mostly physical prior-dependent but are also able to predict photo-z's in a wide
photometric range; they are also capable to provide a chi-square based estimation of the Probability Density
Function (PDF) for all photo-z's.
Empirical methods are instead embedding the information about physical priors, are able to produce accurate
photo-z's only within the photometric ranges imposed by the spectroscopic knowledge base (KB) and
can easily incorporate information, such as the surface brightness of
galaxies, galaxy profiles, concentration,
angular sizes or environmental properties\cite{sadeh2015}. They, however, cannot achieve high accuracy outside the range
defined by the KB and, due to the hidden nature of the flux-redshift correlations, it is quite difficult to derive
reliable PDFs.

\section{The PDF}

From a  statistically rigorous point of view, a PDF is an intrinsic property of a certain phenomenon, regardless the measurement methods that allow to quantify the phenomenon itself.
On the contrary, in the specific case discussed here, the PDF is usually dependent both on the measurement
methods (and chosen internal parameters of the methods themselves) as well as on the physical assumptions
made.
In absence of systematics, factors affecting the PDF are: photometric errors, internal errors of the methods,
statistical biases.
A series of methods have been developed  to derive PDFs, not only for each single source within a catalogue,
but also to estimate the so-called cumulative PDF for a whole sample of galaxies (through the stacking of the
individual PDFs).

ML based regression models look for the mapping between the input parameters and an associated likelihood function spanning the entire redshift region, properly divided in classes (e.g. redshift bins).
Such likelihood is expected to peak in the region where the true redshift actually is, while in the regions where
the uncertainty is high, the same likelihood is expected to be flat.
The purpose is therefore to differentiate between the so-called signal (belonging to a given bin) and the background objects, not belonging to the bin (cf. \cite{sadeh2015,bonnet2013,carrasco2013,carrasco2014a,carrasco2014b}). All these solutions, however, have their limitations as it is discussed in \cite{cavuoti2016a}.

With METAPHOR we try to account in a coherent manner for the uncertainties on the photometric data and to build a Probability Density Function of the error distribution of photometric redshifts predicted by any method by making use of self-adaptive and embedding physical priors techniques to estimate photo-z's.

\section{METAPHOR}

\begin{figure}
\centering

\includegraphics[width=0.45\textwidth]{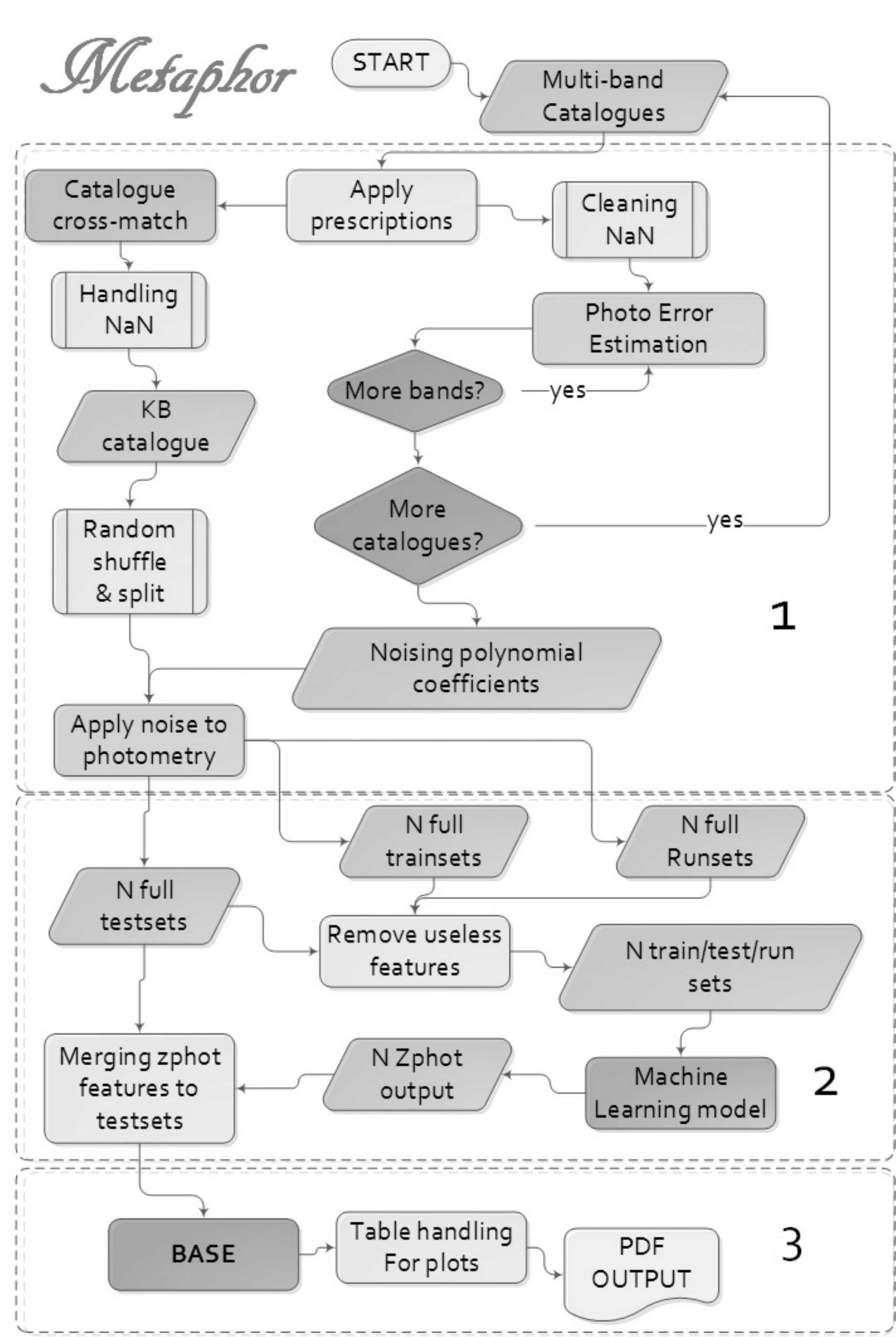}

\caption{Functional block diagram of the METAPHOR workflow.}
\label{fig:base_full}
\end{figure}

METAPHOR (Machine-learning Estimation Tool for Accurate PHOtometric Redshifts) includes all designed
functionalities needed to obtain a PDF for any photo-z prediction experiment done with empirical methods.
In Fig. \ref{fig:base_full} we outline the METAPHOR functional block diagram.
From a logical point of view the method can be divided on the following processing functions:

\begin{enumerate}
 \item Data Pre-processing: photometric evaluation and error estimation of the multi-band catalogue used as KB. This phase includes also the photometry perturbation of the KB;
 \item Photo-z prediction: training/test phase to be performed through the selected empirical method (in this case
 $\mu$MLPQNA, which stands for multi-thread-MLPQNA). It embeds the preliminary catalogue column cutting;
 \item PDF estimation: this phase is related to the BASE method (discussed below) designed and implemented to furnish a PDF evaluation for the photo-z produced.	
\end{enumerate}

In the context of photo-z prediction with empirical supervised methods, a Knowledge Base is a data set composed by objects for which both photometric and spectroscopic information is given.
At the user's convenience, such set should be randomly divided into several sub-sets, with arbitrary splitting percentages, in order to compose, respectively, the training, validation and test datasets.
The training set is used during the learning phase; the validation set can be used to check the training correctness (to avoid overfitting); the third (blind) set of data (test set) is used to statistically evaluate the prediction performance and related error estimate (for instance the PDF of predicted photo-z's).
In our case, the validation set was embedded into the training phase since we used k-fold cross validation \cite{geisser1975}.

From a theoretical point of view, the characterization of photo-z predicted by empirical methods should take into account
the distribution of the photometric errors, the correlation between photometric and spectroscopic errors,
and should allow a good disentanglement of the photometric uncertainty contribution from the one intrinsic to the
method itself. In order to reduce variance, there is also a general agreement to perform the analysis by binning the parameter space.
The right choice of the bin size is however a still unsolved problem due to the risk of information loss induced by
aliasing in the case of high density binning, and by masking in the case of an under-sampling of the parameter space.

In order to deal with the photometric errors, the BASE method introduced in the METAPHOR pipeline starts from a polynomial fitting of the mean error distribution to derive a multiplicative factor for the Gaussian random seeds to be algebraically added to magnitude values.
In order to handle different bands, we allowed for different multiplicative constants
for each band.

The derived perturbing laws are then applied to the cross-matched catalogue.
Given a spectroscopic sample randomly shuffled and split into training and test datasets, METAPHOR perturbs the photometry of the test dataset, in order to obtain an arbitrary number N (statistically consistent) of test sets with a varying quantity of photometric noise contamination.
Then the N+1 test sets (i.e. N noised sets + original not noised) are submitted to the trained model, thus obtaining N estimates of photo-z's. With these values we perform a binning in photo-z and for each one we calculate the probability that a given photo-z value belongs to each bin.
The pseudo-algorithm can therefore be summarised as it follows:

\begin{enumerate}
\item{INPUT: the KB (train + test sets), the photo-z binning step $B$ (default $0.01$) and the spectroscopic redshift (zspec) Region of Interest (RoI) $[Z_{min}, Z_{max}]$ a typical use case is [0,1];}
\item{Produce $N$ photometric perturbations, thus obtaining $N$ additional test sets;}
\item{Perform $1$ training (or $N+1$ trainings) and $N+1$ tests;}
\item{Derive and store: the number of photo-z bins $(Z_{max}-Z_{min})/B$, the $N+1$	photo-z calculated values and the number of photo-z $C_{B,i} \varepsilon [Z_{i}, Z_{i+B}[$;}
\item{Calculate the probability that a photo-z belongs to all given bins: $PDF(Photo-z) = (P(Z_{i} \leq Photo-z < Z_{i+B}) = C_{B,i}/N+1)_{[Z_{min}, Z_{max}]}$;}
\item{Calculate and store statistics.}
\end{enumerate}
{\bf
The results of the photo-z calculations were evaluated using a standard set of statistical
estimators for the quantity $\Delta z = (\zs-\zp)/(1+\zs)$ on the objects in the blind test
set, as listed in the following:

\begin{itemize}
\item bias: defined as the mean value of the residuals $\Delta z$;
\item $\sigma$: the standard deviation of the residuals;
\item $\sigma_{68}$: the radius of the region that includes $68\%$ of the residuals close to 0;
\item $NMAD$: the Normalized Median Absolute Deviation of the residuals, defined as
$NMAD(\Delta z) = 1.48 \times Median (|\Delta z|)$;
\item fraction of outliers with $|\Delta z| > 0.15$;
\item skewness: measurement of the asymmetry of the probability distribution of a
real-valued random variable about its mean.
\end{itemize}

Furthermore, in order to evaluate the cumulative performance of the PDF we computed
the following three estimators on the \textit{stacked} residuals of the PDF's:

\begin{itemize}
\item $f_{0.05}$: the percentage of residuals within $\pm 0.05$;
\item $f_{0.15}$: the percentage of residuals within $\pm 0.15$;
\item $\Braket{\Delta z}$: the weighted average of all the residuals of the \textit{stacked} PDF's.
\end{itemize}
}
\section{An application to the SDSS-DR9 data.}

The Sloan Digital Sky Survey (SDSS, \cite{sdss}), combines multi-band photometry and fiber-based spectroscopy, providing all information required to constrain the fit of a function mapping the photometry into the spectroscopic redshift space. The KB was extracted from the spectroscopic sample of the SDSS-DR9, by collecting objects with specClass \textit{galaxy} together with their photometry ($psfMag$ type magnitudes) and rejecting all objects with non-detected information in any of the five SDSS photometric bands.
From the original query we extracted $\sim 50,000$ objects to be used as train and $\sim 100,000$ objects to be used for the blind test set. Details on the KB can be found in \cite{brescia2014} .

The METAPHOR procedure applies to any empirical photo-z estimation model.
Therefore, we tested it with three different empirical methods: MLPQNA neural network, KNN  and Random Forest, and compared their results with a completely independent method: the \textit{Le Phare} SED template fitting technique.\\

\noindent \textit{MLP with Quasi Newton Algorithm}: the MLPQNA model, i.e. a Multi Layer Perceptron feed-forward neural network trained
by the Quasi Newton learning rule, belongs to the Newton's methods aimed at finding
the stationary point of a function by means of an approximation of the Hessian of
the training error through a cyclic gradient calculation. The implementation makes
use of the known L-BFGS algorithm (Limited memory - Broyden Fletcher Goldfarb Shanno;
\cite{byrd1994}), originally designed to solve optimization problems characterized
by a wide parameter space. The description details of the MLPQNA model have been
already extensively discussed elsewhere (cf. \cite{brescia2013,dame,cavuoti2012,cavuoti2014,cavuoti2014b,cavuoti2015}).\\

\noindent \textit{K-Nearest Neighbor}:
in a KNN (K-Nearest Neighbors; \cite{cover1967}) the input consists of the k
closest training examples in the parameter space. A photo-z is estimated by averaging
the targets of its neighbors. The KNN method is based on the selection of the N training
objects closest to the object currently analyzed. Here closest has to be intended in
terms of Euclidean distance among all photometric features of the objects. Our implementation
makes use of the public library scikit-learn \cite{pedregosa}.\\

\noindent \textit{Random Forest}:
Random Forest (RF; \cite{breiman2001}) is a supervised model which learns by generating
a forest of random decision trees, dynamically built on the base of the variations in the
parameter space of the training sample objects. Each single or group of such trees is
assumed to become representative of specific types of data objects, i.e. the best candidate
to provide the right answer for a sub-set of data similar in the parameter
space \cite{cavuoti2015, hoyle2015}.\\

\noindent \textit{Le Phare SED fitting}.
To test the METAPHOR workflow we used as a benchmark the \textit{Le
Phare} code to perform a SED template fitting experiment
\cite{arnouts1999,ilbert2006}. SDSS observed magnitudes were
matched with those predicted from a set of SEDs. Each SED template
was redshifted in steps of $\Delta z = 0.01$ and convolved with the
five SDSS filter transmission curves.

The following merit function is then minimized:

\begin{equation}
\chi^{2}(z,T,A) = \sum_{i=1}^{N} { \left( \frac{F^{i}_{\rm obs}-A\times F^{i}_{\rm pred}(z,T)}{\sigma^{i}_{obs}} \right)^2},
\end{equation}
\noindent where $F^{i}_{\rm pred}(z,T)$ is the flux predicted for a SED template T at redshift z, $F_{\rm obs}^{i}$ is the observed
flux and $\sigma_{obs}^{i}$ the associated error. $N$ is the number of filters (five in our case). The photometric redshift is determined from the
minimization of $\chi^{2}(z,T,A)$ varying the three free parameters z, T and the normalization factor A.
Details on the specific implementation of \textit{Le-Phare} used here and on how it derives PDFs,
can be found in \cite{cavuoti2016a}.
We wish to stress that we used a basic implementation of the \textit{Le Phare} code, not
taking into account the systematics in the templates, datasets, optimizations
(\cite{brammer2008,Ilbert+09,tanaka2015}), and only imposing a
flat prior on the absolute magnitudes.

\section{Results and discussion}\label{SEC:discussion}

In a previous paper \cite{brescia2014c} we already used the MLPQNA method to derive
photometric redshifts for the SDSS-DR9 obtaining an accuracy better than the one
presented here ($\sigma = 0.023$, $bias \sim 5 \times 10^{-4}$ and $\sim 0.04\%$
of outliers against, respectively, $0.024$, $0.0063$ and  $0.12\%$). This apparent
discrepancy can be easily understood if we take into account that the
spectroscopic KB used in the previous work was much larger than the one used here
(in \cite{brescia2014c} $\sim 150,000$ objects were used for the train and $\sim
348,000$ for the test set while in the present work only $\sim 50,000$ objects were
used for the training phase).

The smaller KB used here is justified by the different
purpose of the present work which aims at assessing the quality of PDF derived by
METAPHOR rather than at deriving a new catalogue of photo-z's for the SDSS-DR9. The
training phase of MLPQNA is in fact computationally intensive and the reduction of
the training sample was imposed by the need to perform a large number of experiments.

As stated above, the use of three different empirical models
(for instance MLPQNA, RF and KNN) has been carried out in order to verify the versatility
of the procedure with respect to the multitude of empirical methods that could be used
to estimate photo-z's. We derived also PDF's with the \textit{Le Phare} method, in order to
evaluate the quality of the produced PDF's using as benchmark a classical SED template fitting
model. In Table~\ref{tab:photozstat} we report the results in terms of the standard set of
statistical estimators used to evaluate the quality of predicted photo-z's for all methods.

The stacked PDF has been obtained by considering bin by bin the average values of the
single PDF's. The cumulative statistics used to evaluate the stacked PDF quality have
been derived by calculating the stacked PDF of the residuals $\Delta z$. In this way,
aside from the evaluation of PDF's for single objects, it is possible to obtain a cumulative evaluation within
the most interesting regions of the error distribution. The related results are shown
in Table~\ref{tab:stackedstat}.

\begin{table}
 \centering
 \begin{tabular}{ccccc}
Estimator	    	& MLPQNA 	& KNN       & RF        & \textit{Le Phare}     	\\ \hline
$bias$		    	& $0.0007$	& $0.0029$  & $0.0035$  &  $0.0009$	\\
$\sigma$	    	& $0.024$	& $0.026$   & $0.025$   &  $0.060$	\\
$\sigma_{68}$   	& $0.018$	& $0.020$   & $0.019$   &  $0.035$     	\\
$NMAD$ 		   	    & $0.017$	& $0.018$   & $0.018$   &  $0.030$     	\\
$skewness$	   	    & $-0.11$	& $0.330$   & $0.015$   &  $-18.076$   	\\
$outliers>0.15$	    & $0.12\%$	& $0.15\%$  & $0.15\%$  &  $0.69\%$    \\ \hline
 \end{tabular}
\caption{Statistics of photo-z estimation performed by the MLPQNA, RF, KNN and \textit{Le Phare} models.} \label{tab:photozstat}
\end{table}

\begin{table}
\centering
 \begin{tabular}{ccccc}
 Estimator		        & MLPQNA 	& KNN        & RF         & \textit{Le Phare}   \\ \hline
 $f_{0.05}$		        & $92.9\%$	& $92.0\%$   & $92.1\%$   & $71.2\%$   \\
 $f_{0.15}$		        & $99.8\%$	& $99.8\%$   & $99.7\%$   & $99.1\%$   \\
 $\Braket{\Delta z}$	& $-0.0011$	& $-0.0018$  & $-0.0016$  & $0.0131$   \\ \hline
 \end{tabular}
\caption{Statistics of the \textit{stacked} PDF obtained by \textit{Le Phare} and by the three empirical models MLPQNA, KNN and RF through METAPHOR.} \label{tab:stackedstat}
\end{table}

\begin{figure*}
\centering
\includegraphics[width=0.4 \textwidth]{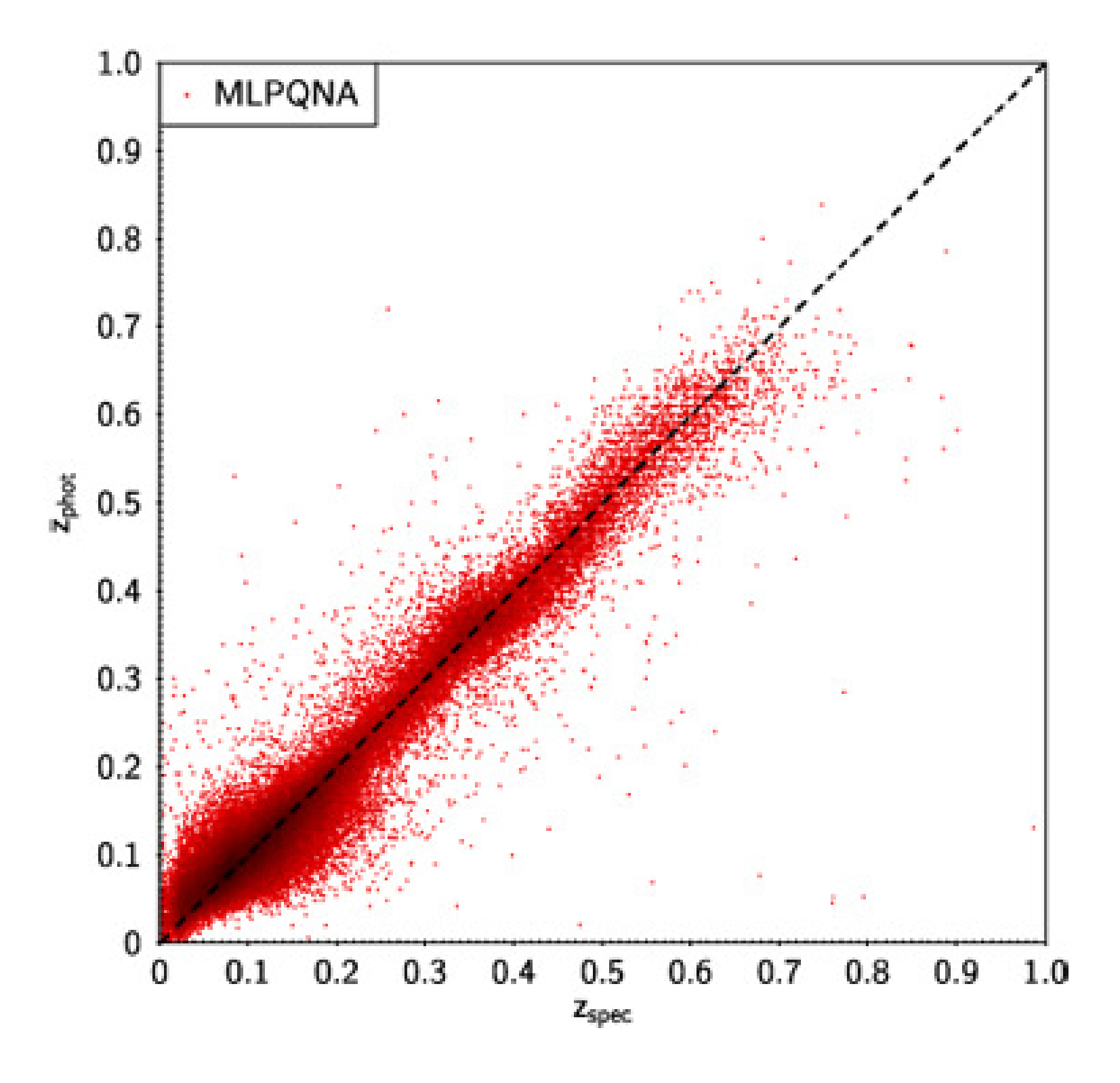}
\includegraphics[width=0.4 \textwidth]{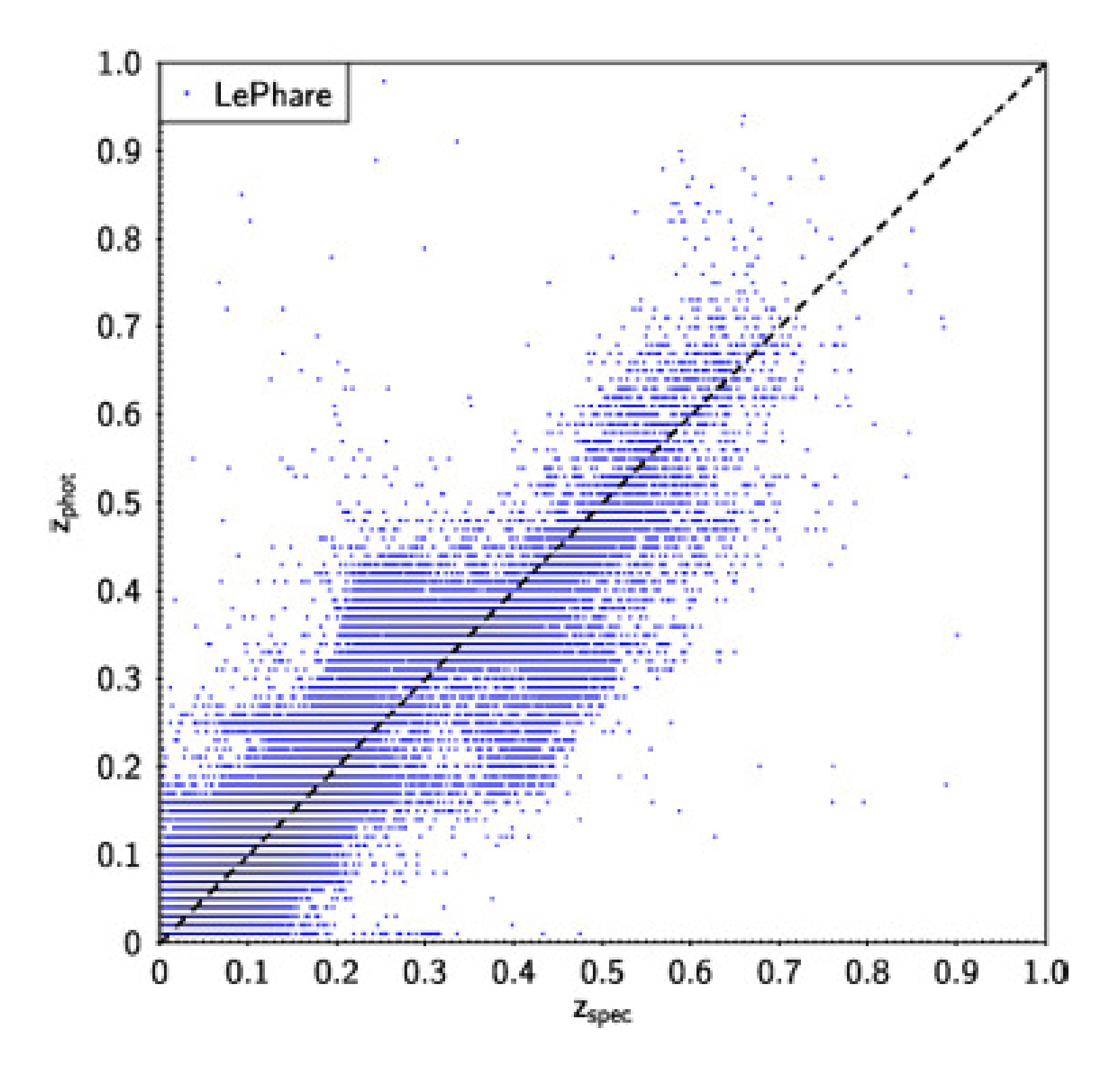}\\
\includegraphics[width=0.4 \textwidth]{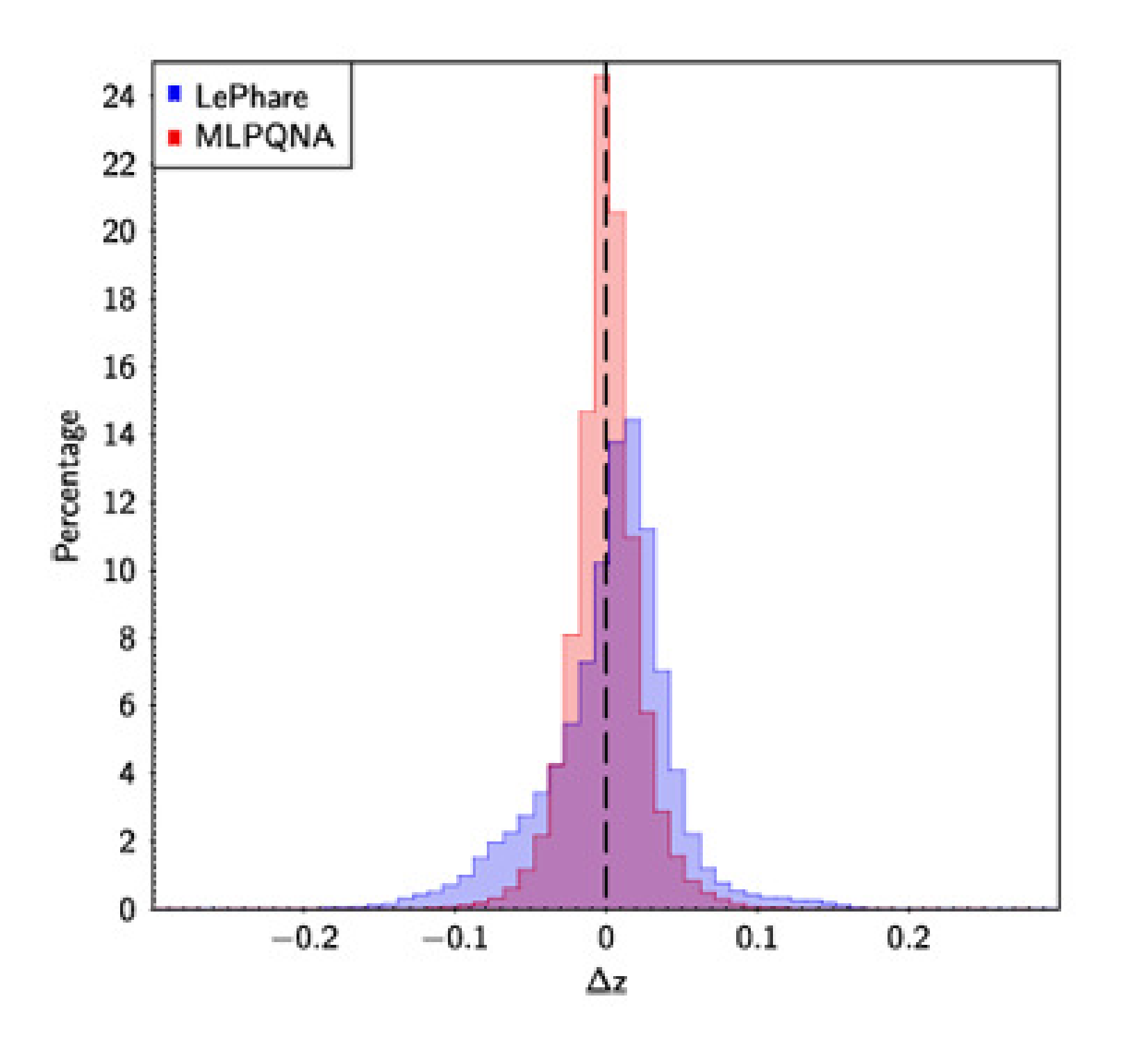}
\includegraphics[width=0.38 \textwidth]{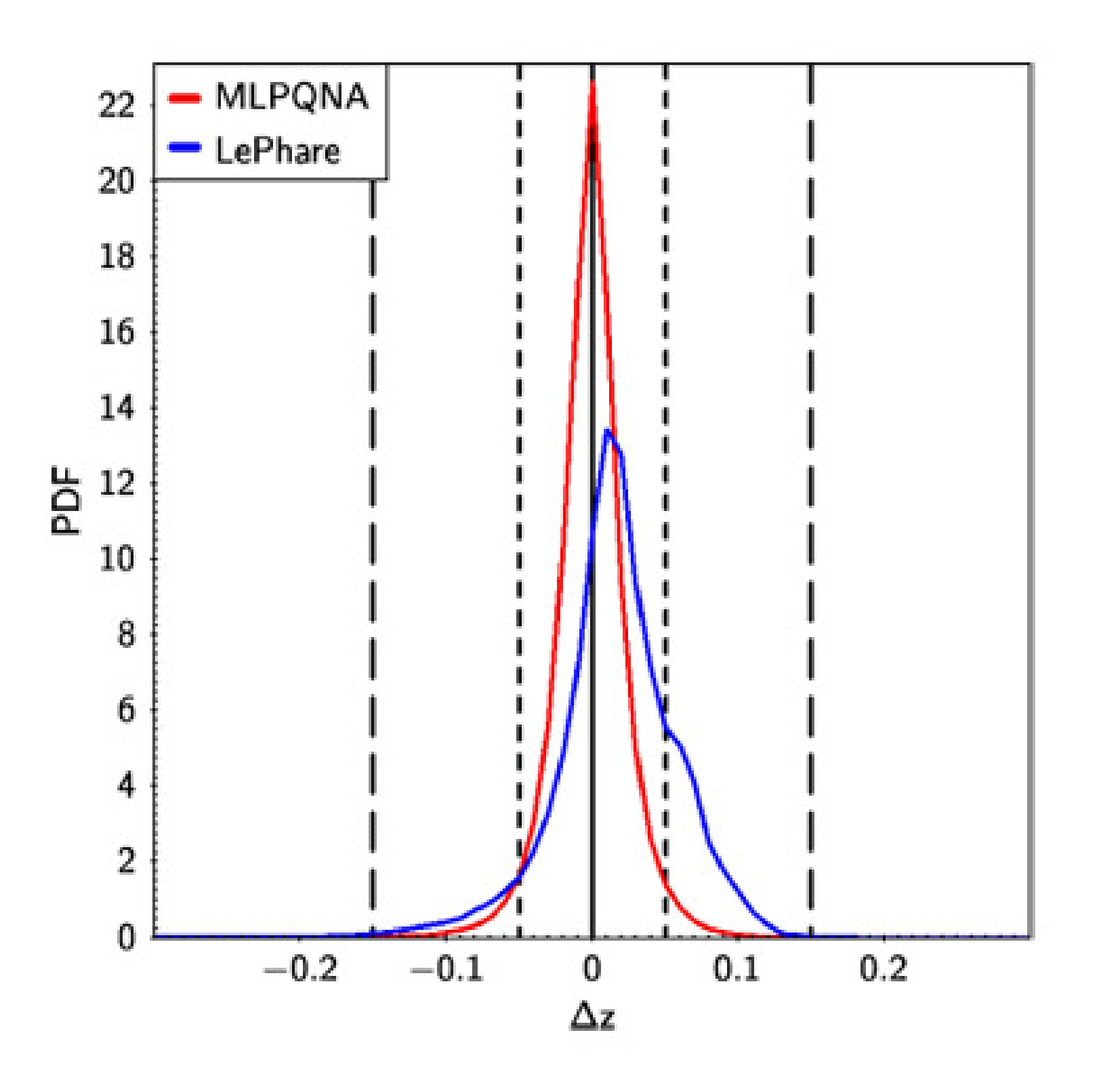}
\caption{Comparison between MLPQNA and \textit{Le Phare}. Left panel of upper row: scatter plot of photometric redshifts as function of spectroscopic redshifts (zspec vs zphot) obtained by, respectively, MLPQNA (left panel) and LePhare (right panel); left panel of lower row: histograms of residuals ($\Delta z$); right panel of lower row: \textit{stacked} representation of residuals of the PDF's (the redshift binning is $0.01$).} \label{fig:scatter}
\end{figure*}

Although there is a large difference in terms of statistical estimators between
\textit{Le Phare} and MLPQNA, as it can be seen from Table~\ref{tab:photozstat}
and Figure~\ref{fig:scatter}, the results of the PDF's in terms
of $f_{0.15}$ are comparable (see Table \ref{tab:stackedstat} and the right panel in
the lower row of Figure~\ref{fig:scatter}). But the greater efficiency of MLPQNA
induces an improvement in the range within $f_{0.05}$, where we find $\sim 92\%$
of the objects against $\sim 72\%$ for \textit{Le Phare}.
Both individual and \textit{stacked} PDF's are more symmetric in the case of the interpolative
methods presented here than for \textit{Le Phare}.
This is particularly apparent from the skewness (see Table \ref{tab:stackedstat}),
which is $\sim 240$ times greater for SED template fitting method; this can be also seen
by looking at the panels in the lower row of Figure~\ref{fig:scatter}.

\begin{table*}
 \centering
 \begin{tabular}{ccccccccc}
Estimator	&	Overall	&	]0, 0.1]&	]0.1, 0.2]	&	]0.2, 0.3]&	]0.3, 0.4]&	]0.4, 0.5]&	]0.5, 0.6]&	]0.6, 1]\\\hline
$bias$		&	-0.0007	&	-0.0005	&	-0.0008		&	-0.0008	&	-0.0016	&	0.0011	&	-0.0022	&	-0.0032	\\
$\sigma$	&	0.024	&	0.022	&	0.024		&	0.030	&	0.026	&	0.028	&	0.033	&	0.037	\\
$\sigma_{68}$	&	0.018	&	0.018	&	0.019		&	0.018	&	0.019	&	0.019	&	0.021	&	0.029	\\
$NMAD$		&	0.017	&	0.016	&	0.017		&	0.016	&	0.017	&	0.017	&	0.020	&	0.027	\\
$skewness$	&	-0.11	&	1.40	&	-0.070		&	-1.04	&	-1.53	&	-2.37	&	-2.47	&	-1.70	\\
$outliers>0.15$	&	0.12\%	&	0.00\% &	0.03\%		&	0.03\%	&	0.29\%	&	0.51\%	&	0.82\%	&	0.35\%	\\\hline
 \end{tabular}
\caption{Tomographic analysis of photo-z estimation performed by the MLPQNA. } \label{tab:photozstatTomo}
\end{table*}

\begin{table*}
\centering
 \begin{tabular}{ccccccccc}
Estimator		&	Overall	&	]0, 0.1]&	]0.1, 0.2]	&	]0.2, 0.3]&	]0.3, 0.4]&	]0.4, 0.5]&	]0.5, 0.6]&	]0.6, 1]\\\hline
$f_{0.05}$		&	92.9\%	&	94.6\%	&	92.4\%	&	90.7\%	&	92.5\%	&	90.0\%	&	84.7\%	&	77.6\%	\\
$f_{0.15}$		&	99.8\%	&	99.9\%	&	99.9\%	&	99.1\%	&	99.6\%	&	99.5\%	&	99.1\%	&	99.2\%	\\
$\Braket{\Delta z}$	&	-0.0011	&	-0.0010	&	-0.0012	&	-0.0009	&	-0.0014	&	-0.0017	&	-0.0017	&	-0.0014	\\\hline
 \end{tabular}
\caption{Tomographic analysis of PDF obtained by MLPQNA. Statistics of the \textit{stacked} PDF obtained by MLPQNA. } \label{tab:stackedstatTomo}
\end{table*}

The model KNN performs slightly worse than MLPQNA in terms of $\sigma$ and $outliers$
rate (Table~\ref{tab:photozstat}), while RF obtains results placed between
KNN and MLPQNA in terms of statistical performance.
The higher accuracy of MLPQNA causes a better
performance of PDF's in terms of $f_{0.05}$, which describes the inner region of the PDF.
However, also in the case of KNN and RF, METAPHOR is capable to produce reliable PDF's,
comparable with those produced for MLPQNA (see Table~\ref{tab:stackedstat}).

\begin{figure*}
\centering
\includegraphics[width=0.3 \textwidth]{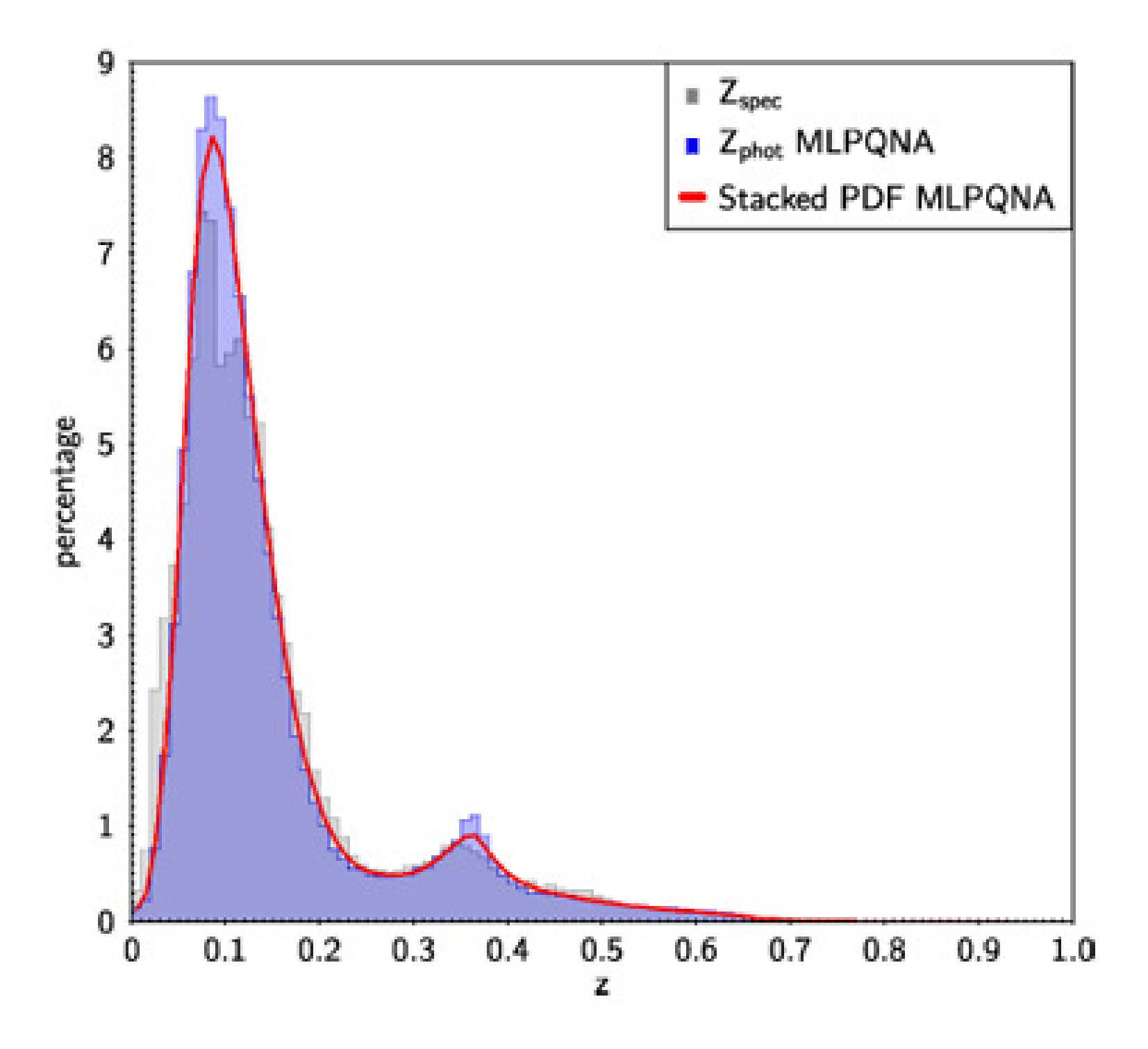}
\includegraphics[width=0.3 \textwidth]{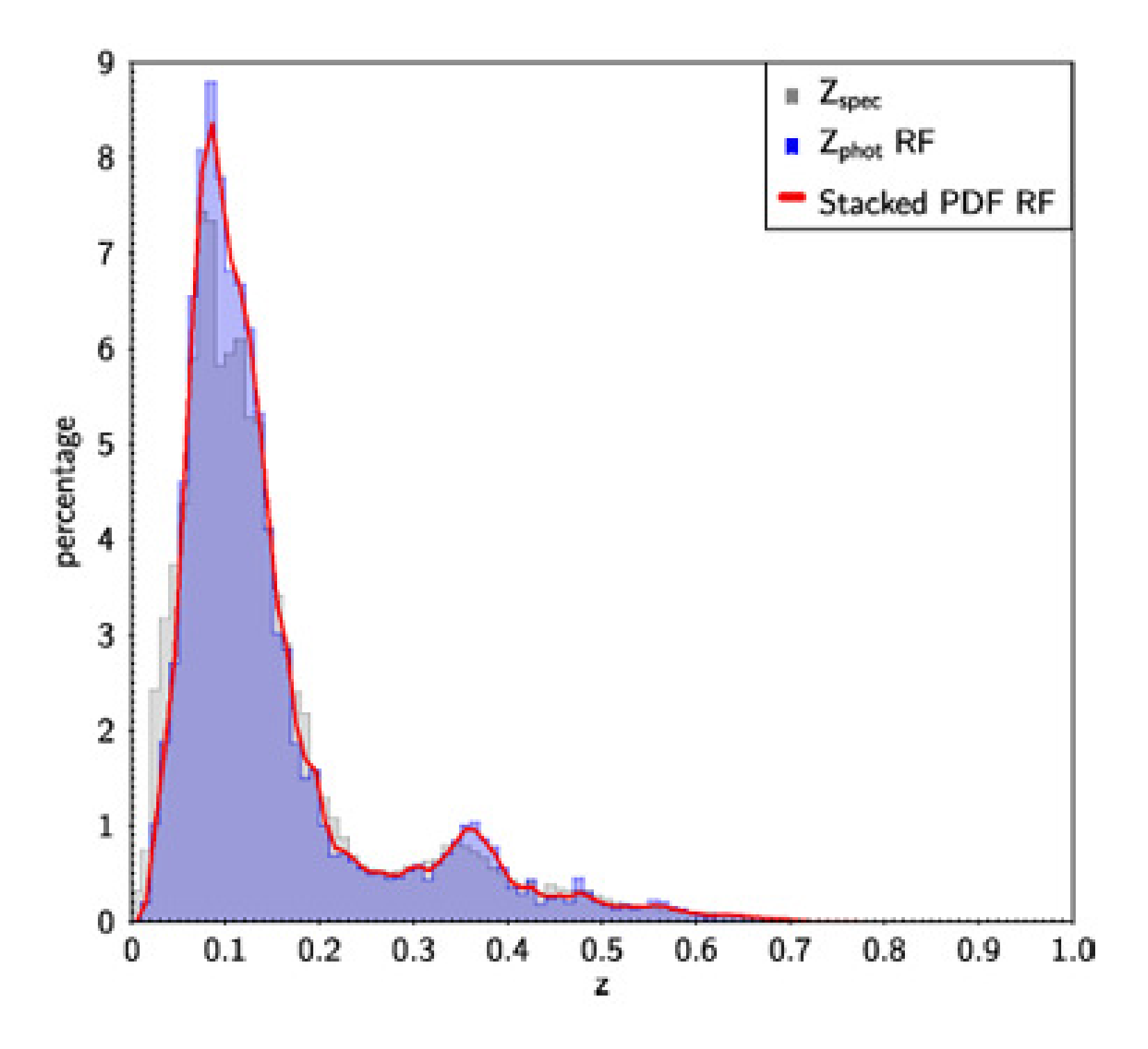}
\includegraphics[width=0.3 \textwidth]{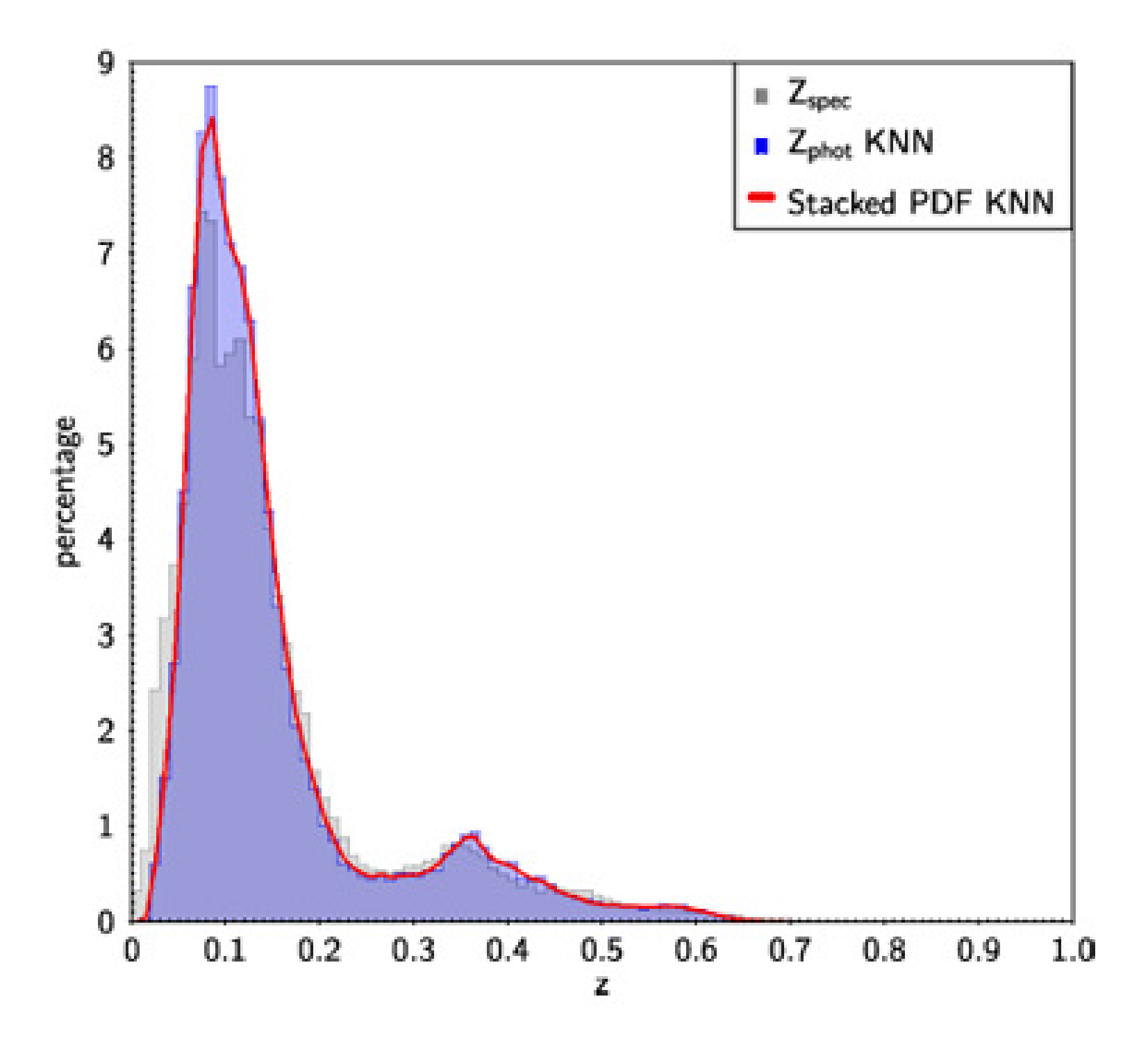}
\caption{Superposition of the \textit{stacked} PDF (red) and estimated photo-z (blue) distributions obtained by METAPHOR with, respectively, MLPQNA, RF, and KNN, on the zspec distribution (in gray) of the blind test set.} \label{fig:stackedpdfdistrib}
\end{figure*}

The efficiency of the METAPHOR with the three empirical methods is made apparent
by looking at the Figure~\ref{fig:stackedpdfdistrib}, where we show the \textit{stacked}
PDF and the estimated photo-z distributions, obtained by METAPHOR with each of the three
models, superposed over the distribution of spectroscopic redshifts. The \textit{stacked}
distribution of PDF's, derived with the three empirical methods, results almost indistinguishable
from the distribution of spectroscopic redshifts, with the exception of two regions: one in
the peak of the distribution at around $z\simeq0.1$ and the other at $z\simeq0.4$. The first one can be understood in terms of a mild overfitting induced by the uneven distribution of objects in the training set.
The second one ($z\simeq0.4$) can be explained by the fact that the break at $4000$ \AA\ enters in the r band at this redshift. It induces  an edge effect in the parameter space which leads our methods to generate predictions biased away from the edges.
However, biases in color-space (averaging over/between degeneracies) specific to the SDSS filters clearly play a role as well.
This confirms the capability of METAPHOR to
work efficiently with different empirical methods regardless of their nature; even a very simple
empirical model like KNN is able to produce high quality PDF's.

By analyzing the relation between the spectroscopic redshift and the PDF's that we produce,
we find that about $22\%$ of zspec falls in the bin of the peak of the PDF but we emphasize that a further $32\%$ of zspec falls one bin far from the peak (in our exercise this means
a distance of 0.01 from the peak), while $35\%$ of the zspec falls outside the PDF.
We analyze in a tomographic way the results in order to verify whether there is different behavior in different regions.
This has been done by cutting the output in bins of zphot (the best guess of our method) and deriving the whole statistics bin by bin. Results are shown in tables \ref{tab:photozstatTomo}, \ref{tab:stackedstatTomo}.

\section{Conclusions}\label{SEC:conclusion}

In this work we introduce METAPHOR (Machine-learning Estimation Tool for Accurate
PHOtometric Redshifts), a method designed to provide a reliable PDF of the error
distribution of photometric redshifts predicted by empirical methods. The method is
implemented as a modular workflow, whose internal engine for photo-z estimation is
based on the MLPQNA neural network (Multi Layer Perceptron with Quasi Newton learning rule).
The METAPHOR procedure can however be applied by making use of any arbitrary empirical
photo-z estimation model. One of the most important goals of this work was to verify the universality
of the procedure with respect to different empirical models. For this reason we
experimented the METAPHOR processing flow on three alternative empirical methods.
Besides the canonical choice of MLPQNA, a powerful neural network that we developed
and tested on many photo-z estimation experiments, the alternative models selected
were Random Forest and the K-Nearest Neighbor. In particular, the choice of KNN has
been mainly driven by taking into account its extreme simplicity with respect to the
wide family of empirical techniques. We tested the METAPHOR strategy and the photo-z
estimation models on a sample of the SDSS DR9 public galaxy catalogue.

The presented photo-z estimation results and the statistical performance of the cumulative
PDF's, achieved by MLPQNA, RF and KNN through the proposed procedure, demonstrate the validity
and reliability of the METAPHOR strategy, despite its simplicity, as well as its general
applicability to any other empirical method.

\section*{Acknowledgments}
MB and SC acknowledge financial contribution from the agreement ASI/INAF I/023/12/1.
MB acknowledges the PRIN-INAF 2014 \textit{Glittering kaleidoscopes in the sky: the
multifaceted nature and role of Galaxy Clusters}.
CT is supported through an NWO-VICI grant (project number $639.043.308$).

\end{document}